\documentclass[aps,pre,floats, twocolumn,showpacs,superscriptaddress]{revtex4-1}

\usepackage{graphicx,epsfig}
\usepackage{graphics,dcolumn,bm,epic, eepic,fleqn,float}
\usepackage{amssymb,amsmath,amsfonts,multirow,rotate,color}
\usepackage{soul}
\usepackage{wasysym}
\usepackage{tabularx}
\usepackage{xcolor}
\usepackage{amsmath, amssymb}

\begin{document}
\title{Homophily and missing links in citation networks}

\author{Valerio Ciotti}
\affiliation{School of Business and Management, Queen Mary University of
  London, Mile End Road, E1 4NS London, UK}
\affiliation{School of Mathematical Sciences, Queen Mary University of
London, Mile End Road, E1 4NS London, UK}

\author{Moreno Bonaventura}
\affiliation{School of Business and Management, Queen Mary University of
  London, Mile End Road, E1 4NS London, UK}
\affiliation{School of Mathematical Sciences, Queen Mary University of
London, Mile End Road, E1 4NS London, UK}

\author{Vincenzo Nicosia}
\affiliation{School of Mathematical Sciences, Queen Mary University of
London, Mile End Road, E1 4NS London, UK}

\author{Pietro Panzarasa}
\affiliation{School of Business and Management, Queen Mary University of
  London, Mile End Road, E1 4NS London, UK}

\author{Vito Latora}
\affiliation{School of Mathematical Sciences, Queen Mary University of
London, Mile End Road, E1 4NS London, UK}
\affiliation{Dipartimento di Fisica e Astronomia, Universit\'a di
  Catania \& INFN, Via S. Sofia, I-95123 Catania, IT}

\begin{abstract} 
Citation networks have been widely used to study the evolution of
science through the lenses of the underlying patterns of knowledge
flows among academic papers, authors, research sub-fields, and
scientific journals. Here we focus on citation networks to cast light
on the salience of homophily, namely the principle that similarity
breeds connection, for knowledge transfer between papers. To this end,
we assess the degree to which citations tend to occur between papers
that are concerned with seemingly related topics or research
problems. Drawing on a large data set of articles published in the
journals of the American Physical Society between 1893 and 2009, we
propose a novel method for measuring the similarity between articles
through the statistical validation of the overlap between their
bibliographies. Results suggest that the probability of a citation
made by one article to another is indeed an increasing function of the
similarity between the two articles. Our study also enables us to
uncover missing citations between pairs of highly related articles,
and may thus help identify barriers to effective knowledge flows. By
quantifying the proportion of missing citations, we conduct a
comparative assessment of distinct journals and research sub-fields in
terms of their ability to facilitate or impede the dissemination of
knowledge. Findings indicate that Electromagnetism and
Interdisciplinary Physics are the two sub-fields in physics with the
smallest percentage of missing citations. Moreover, knowledge transfer
seems to be more effectively facilitated by journals of wide
visibility, such as Physical Review Letters, than by lower-impact
ones. Our study has important implications for authors, editors and
reviewers of scientific journals, as well as public preprint
repositories, as it provides a procedure for recommending relevant yet
missing references and properly integrating bibliographies of papers.
\end{abstract}

\maketitle

\section{Introduction}

Among the broad category of information networks, including the Word
Wide Web \cite{barabasi1999emergence}, email exchange networks
\cite{milo2002network}, and phone call networks
\cite{eagle2009inferring}, the networks of citations between academic
papers have been widely investigated to uncover patterns and dynamics
of knowledge transfer, sharing, and creation in science
\cite{larsen2010rate,leicht2007structure,price1965citation,Price}. The
nodes of citation networks are academic papers, each containing a
bibliography with references to previously published work. Typically,
a directed link is established from one paper to another if the former
cites the latter in its bibliography. Because papers can only cite
other papers that have already been published, all directed links in
citation networks necessarily point backward in time. Citation
networks are therefore {\em directed acyclic graphs}, i.e., they do
not contain any closed loops of directed links \cite{newman}.

Since the seminal work by Derek de Solla Price on the distribution of
citations received by scientific articles
\cite{price1965citation,Price}, citation networks have extensively
been studied to shed light on the mechanisms underpinning the
evolution, diffusion, recombination, and sharing of knowledge over
time \cite{goldberg2014modelling,calero2008combining}. The reason why
citation networks are crucial to understanding and modelling
scientific production is clear. Although citations can serve different
functions -- for instance, they acknowledge the relevance of previous
work, they help the reader of a paper to gather additional information
about a specific topic, they point to related work or, sometimes, they
can also express disagreement with, or level criticism against, a
position endorsed in a paper \cite{catalini2015incidence} -- the
number of citations received is generally regarded as an indication of
the relevance and quality of a paper as well as of its authors'
prestige and scientific success \cite{fortunato}. Certainly, citation
networks can be used to reconstruct the communication flows among
different scientific communities and infer the relation among
different research topics and sub-fields
\cite{sinatra2015century}. Recent work on citation networks has indeed
proposed a new method for highlighting the role of citations as
conduits of knowledge. For instance, Clough et
al. ~\cite{clough2015transitive,clough2014dimension} have proposed
reduction methods to filter out the relevant citations preserving the
causal structure of the underlying network and of knowledge flows.

In this paper, we study citations from a different perspective. First,
we assess the extent to which the occurrence of a citation between two
papers is driven by the similarity between them. Specifically, we
investigate empirically a large data set of articles published in the
journals of the American Physical Society (APS) \cite{aps}, and we
measure the similarity between any two articles by drawing on, and
extending, a method originally proposed by Tumminello et al. in
Ref.~\cite{tumminello2011statistically,tumminello2} that enables us to
statistically validate the overlap between the bibliographies of the
two articles. Results suggest that the number citations made by one
article to another is indeed an increasing function of the similarity
between the two articles. Our findings thus indicate that the creation
of links in citation networks can be seen as governed by {\em
  homophily}, namely the principle that similarity breeds connection
\cite{aral,kossinets,lazerfeld,mcpherson2001birds}.

Second, we propose a novel method for identifying missing links in
citation networks. The gist of our argument is simple. We focus on
pairs of articles characterised by high degrees of similarity; if a
citation between them is missing, we regard the lack of a directed
link as a signature of a relevant yet unrecorded flow of knowledge in
the network. By uncovering pairs of published articles with missing
citations, we rank the APS journals and topics according to the
incidence of missing data on knowledge flows.

Our method has important implications for the analysis not only of
published articles, but also of newly posted preprints on online
archives, or of manuscripts submitted to scientific
journals. Specifically, our method can be used to suggest interesting
work and relevant literature that could, in principle, be included in
the bibliography of recently posted or submitted preprints. As we
witness a continuously increasing production of preprints and
publication of new articles, it has become particularly difficult for
authors to keep abreast of scientific developments and relevant works
related to the domain of interest. As a result, lack of knowledge of
prior or current related work and missing relevant citations may occur
quite often. The method presented in this paper can help the
scientific community precisely to address this problem. In particular,
it can be used not only by authors to integrate the bibliographies of
their work, but also by editors of scientific journals to uncover
missing citations and identify the appropriate reviewers for the
papers they are considering for publication.

The paper is organised as follows. In Section~\ref{Quantifying article
  similarity}, we introduce and discuss our method for evaluating
similarity between articles based on the statistical significance of
the overlap between their respective bibliographies. In
Section~\ref{Results}, we apply our method to all articles published
in the journals of the APS. We show that citations between articles
are positively correlated with their similarity, and we then identify
missing links between similar articles published in different fields
and in different journals. In Section~\ref{Conclusions}, we summarise
our findings and discuss implications, limitations, and avenues for
future work. Finally, in Section~\ref{Materials and Methods}, we
describe the data set and the validation technique used in our
analysis.

\section{Quantifying similarity between articles}
\label{Quantifying article similarity}

Similarity between two articles can be measured in a number of ways. A
straightforward, yet labour-intensive way of comparing articles is to
semantically analyse their entire texts. Alternatively, similarity can
be simply based on the co-occurence of a few relevant concepts or
keywords in the titles or abstracts of the articles. Moreover,
similarity can be measured through the co-occurrence of classification
codes, such as those included in the Physics and Astronomy
Classification Scheme (PACS), which help identify the research areas
to which each article belongs \cite{schummer}. Here, we propose an
alternative measure of similarity based on the comparison between the
bibliographic lists of references included in two articles. Our
hypothesis is that, if two articles are concerned with related aspects
of the same discipline or research problem, then their bibliographies
will exhibit a substantial overlap. We shall therefore introduce a
method for assessing the statistical significance of the overlap
between the lists of references of two articles, and we shall then use
the statistically validated overlap as as measure of the similarity
between the two articles.

\subsection{Overlap between reference lists as a measure of similarity between articles}

\begin{figure*}
  \begin{center}
    \includegraphics[width=6in]{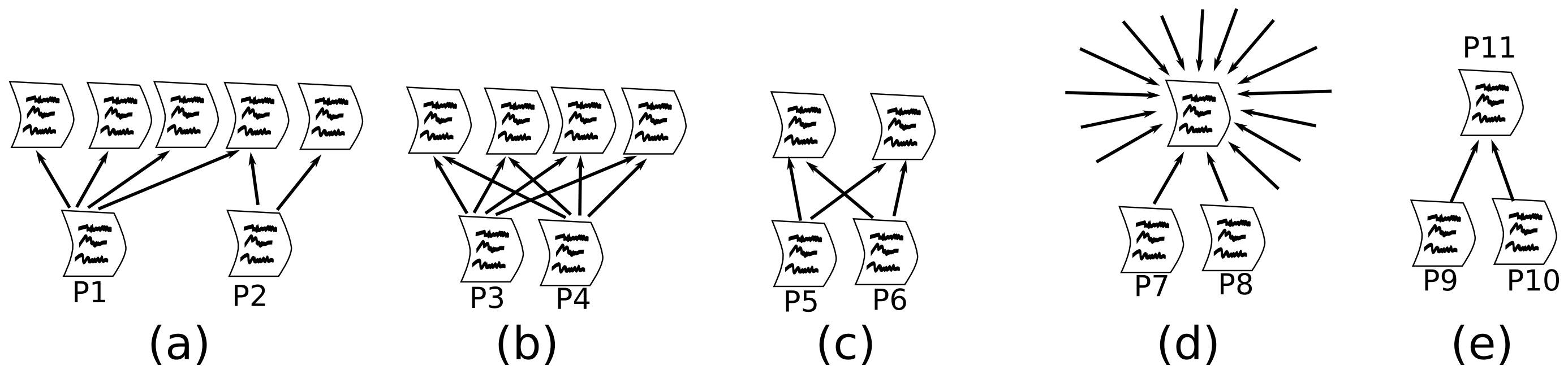}
  \end{center}
  \caption{\textbf{Quantifying the similarity between two articles
      based on their bibliographies.}  The similarity between two
    articles can be defined in terms of the overlap between their
    reference lists. The two articles P1 and P2 in panel (a) share
    only one citation; they should therefore be considered less
    similar than articles P3 and P4 in panel (b) which share four
    citations.  This difference can be captured by the Jaccard index,
    which is equal to $0.2$ in the former case and to $1.0$ in the
    latter. However, the Jaccard index is equal to $1.0$ also for the
    two articles in panel (c), which instead share only two
    citations. If citations are interpreted as proxies for knowledge
    flows, then the similarity between article P7 and P8 in panel (d),
    which cite a highly-cited article, should be smaller than the
    similarity between articles P9 and P10 in panel (e), which instead
    are the only two articles citing P11. Our similarity measure,
    based on statistical validation, properly takes these
    heterogeneities into account.}
  \label{fig:fig1}
\end{figure*}

A natural way to quantify the overlap between two given sets $Q_i$ and
$Q_j$ is the Jaccard index, which is defined as the ratio between the
number of common elements in the two sets and the total number of
elements in the union of the two sets:
\begin{equation}
  J_{ij} = \frac{ |Q_i \cap Q_j|}{|Q_i \cup  Q_j|}.
\label{eq:jacard}
\end{equation}
Notice that, in general, if two sets share a higher number of
elements, then their Jaccard index will increase, and in particular
$J_{ij} = 1$ only if $Q_i \equiv Q_j$, while $J_{ij} = 0$ if the two
sets do not share any element. An example of the suitability of the
Jaccard index for measuring the similarity between the bibliographies
of two papers is provided in Fig.~\ref{fig:fig1}(a)-(b). Here the two
sets $Q_i$ and $Q_j$ represent, respectively, the articles in the two
reference lists of the two articles $i$ and $j$.  Since article P1 and
article P2 share only one reference over a total of five, their
Jaccard index is equal to $0.2$. Conversely, the two articles P3 and
P4 in panel (b) have a Jaccard index equal to $1.0$, since the overlap
between their reference lists is complete.

However, the use of the Jaccard index has some drawbacks.  First, the
value of $J_{ij}$ is always bounded from above by $\frac{\min(|Q_i|,
  |Q_j|)}{|Q_i| + |Q_j|}$. This means that if the sizes of the two
sets are remarkably different, their similarity is primarily
determined by the size of the smallest of the two sets. As a
consequence, large sets tend to be characterised by relatively small
values of similarities with other smaller sets.  In addition to this,
the Jaccard index does not distinguish between pairs of identical sets
having different sizes. In particular, if we consider two identical
sets $(Q_i, Q_j)$ of size $N_1$ and two other identical sets $(Q_m,
Q_n)$ of size $N_2$, then we have $J_{ij} = J_{mn} = 1$, regardless of
the values of their sizes $N_1$ and $N_2$. For instance, the Jaccard
index of articles P5 and P6 is equal to $1.0$ and is identical to that
of articles P3 and P4, even though P3 and P4 share a larger number of
references. In the case of bibliographic references, this degeneracy
of the Jaccard index is very important. In fact, if we interpret
references as proxies for knowledge flows from cited to citing
articles, then it would be reasonable to associate a higher value of
similarity to a pair of articles that share a large number of
references than to a pair sharing only few references, since the
former pair is expected to draw on a more similar scientific
background. In particular, we would expect the two articles in panel
(b) to be assigned a value of similarity larger than the two articles
in panel (c).

Another drawback of a bare count of the number of common references is
that some citations can, in principle, be more important than
others. Consider the two cases depicted in
Fig.~\ref{fig:fig1}(d)-(e). In panel (d), articles P7 and P8 have an
identical set of references, consisting in the citation of a single
highly-cited article. Also in panel (e), both articles P9 and P10 cite
the same article.  However, in this case the cited article does not
receive any citation from other articles. Now, since our aim is to
quantify the similarity between articles, a citation to a highly-cited
paper, such as a review article, should be considered less relevant
than a citation to a more specialised or less visible article, which
is cited only by articles concerned with a certain specific topic. In
other words, it would be preferable to associate a higher relevance to
the single citation shared by articles P9 and P10 in
Fig.~\ref{fig:fig1}(e) than to the citation to other highly cited
articles shared by articles P7 and P8 in Fig.~\ref{fig:fig1}(d), and
thus to conclude that articles P9 and P10 are more similar than
article P7 and P8.

\subsection{Defining statistically significant bibliographic overlaps}

The method we propose here allows us to overcome the drawbacks of the
Jaccard index discussed above and illustrated in Fig.~\ref{fig:fig1}.
The method is based on an extension of the so-called {\em
  Statistically Validated Network (SVN)} approach to the case of
directed unipartite graphs. Statistically Validated Networks were
introduced by Tumminello et
al.~\cite{tumminello2011statistically,tumminello2} as a method to
filter out statistically irrelevant information from bipartite graphs,
such as user-item networks deriving from purchase systems or product
reviews. In such systems, a set $A$ of nodes (e.g., buyers, users)
express preferences over another set $B$ of nodes (e.g., books,
movies, services).  Those preferences or selections are represented by
directed links from nodes in set $A$ to nodes in set $B$. The idea
behind SVNs is that the similarity between two nodes $i$ and $j$ in
the set $A$ can be expressed in terms of the co-occurrence of their
selections of nodes in $B$, and in particular that it is possible to
attach a statistical significance, namely a $p$-value, to each set of
common selections made by $i$ and $j$.

Citation networks are not bipartite graphs. They are also different
from user-item networks because each article in general can only cite
other articles that have already been published, and can only receive
citations from other articles that will be published after its
publication date. Nevertheless, it is possible to draw upon the same
idea used to construct bipartite statistically validated networks, and
define a similarity between two articles based on the overlap between
their reference lists.

Let us consider two sets of nodes, $A$ and $B$. The set $A$ contains
all the articles with more than zero outgoing citations, $A = \{ i \in
V \, | \, k_i^{\rm out}>0\}$, while the set $B$ contains all the
articles that have received at least two citations, $B = \{ i \in V \,
| \, k_i^{\rm in}>1\}$. It is worth noticing that $A\cap B \neq
\emptyset$, i.e., the two sets may share some articles, since in
general each article cites and is cited by other articles. We denote
by $N_A = |A|$ and $N_B=|B|$ the cardinality of the two sets.  The
method associates a statistical significance to the similarity between
a pair of nodes $(i,j)$ in $A$ by comparing the number of
co-occurrences of citations in their reference lists against the null
hypothesis of random co-occurrence of citations to one or more
articles in $B$. In this way, the method allows us to identify pairs
of nodes in $A$ characterised by overlaps between citations to
elements in $B$ which are statistically different from those expected
in the null model.

The method works as follows. For each value $k$ of in-degree observed
in the citation network, we consider the set of nodes $S^k = S^k_B
\cup S^k_A$, where $S^k_B \subset B $ contains all $N_B^k = |S^k_B|$
articles with in-degree equal to $k$, and $S^k_A \subset A$ contains
all articles that cite at least one element in $S^k_B$. Notice that
the set $S^k$ is, by construction, homogeneous with respect to the
in-degree of the elements belonging to the set $B$. Then, for each
pair of articles $i,j\in S_A^{k}$, we indicate by $d_i$ and $d_j$
their respective number of citations directed towards the elements of
$S^k_B$. Under the hypothesis that the articles $i$ and $j$ cite,
respectively, $d_i$ and $d_j$ distinct elements uniformly at random
from $S^k_B$, the probability that they select the same $X$ articles
is given by the hypergeometric probability function:

\begin{equation}
\mathcal{P}(X \,| \,N_B^k,d_i ,d_j) = \frac{{{d_i}\choose{X}} {{N_B^k-d_i}\choose{d_j - X}}}{{{N_B^k}\choose{d_j}}}.
\label{hyper}
\end{equation} 
Thus, we can associate a $p$-value to each pair of nodes $i,j\in S_A^{k}$:

\begin{equation}
q_{ij}(k) = 1 - \sum_{X=0}^{N_{ij}^k -1} \mathcal{P}(X \, | \,
N_B^k,d_i,d_j),
\label{eq:p_value}
\end{equation}
\noindent where $N_{ij}^k$ is the measured number of references that
$i$ and $j$ have in common in the set $S^k_B$. The $p$-value,
$q_{ij}(k)$, is therefore the probability that the number of articles
in the set $S^k_B$ that both $i$ and $j$ happen to jointly cite by
chance is $N_{ij}^k$ or more. We repeat the procedure for all possible
values of in-degree $k$ from $k_{\rm min}$ to $k_{\rm max}$, so that
each pair of articles $(i,j)$ is, in general, associated with several
$p$-values, one for each value of in-degree $k$ of the articles in
their reference lists. Once all the $p$-values have been computed, we
set a significance threshold $p^*$ and validate all the pairs of nodes
that are associated with a $p$-value smaller than the threshold
$p^*$. Given a value of the statistical threshold, only the validated
pairs of articles are considered similar at that significance level.

However, because each pair of articles $(i,j)$ can be associated with
multiple $p$-values, it is necessary to perform hypothesis-testing
multiple times. In this case, if we choose a confidence level or
significance threshold $p^*$, say $1\%$ confidence level ($p^*=0.01$),
the various $p$-values associated with the same pair of nodes are not
compared directly with the chosen significance threshold $p^*$, but
with a rescaled threshold that appropriately takes the number of tests
performed into account. As a method for multiple testing we use the
False Discovery Rate
(FDR)~\cite{tumminello2011statistically,Benjamini} (see
Section~\ref{Materials and Methods} for details). Ultimately, we
identify the set $\mathcal{M}(p^*)$ of all pairs of nodes whose
similarity is statistically significant at the confidence threshold
$p^*$. In what follows, we shall denote by
$M(p^{*})=\left|\mathcal{M}(p^*)\right|$ the cardinality of such set.
In principle, since each pair of articles $(i,j)$ can belong to
different sets $S^k$ (and, as a result, can be associated with several
$p$-values $q_{ij}(k)$), it would be possible to define a similarity
weight $w_{ij}(p^*)$ for each pair $(i,j)$ as the number of times that
the pair is validated at the confidence threshold $p^*$. In other
words, $w_{ij}(p^*)$ would be the number of sets $S^{k}$ for which
$q_{ij}(k)$ passes the statistical test. However, we do not consider
this possibility here, but simply assume that a pair of articles
$(i,j)$ belongs to the set $\mathcal{M}(p^*)$ if at least one of the
$p$-values $q_{ij}(k)$ passes the statistical test at the confidence
threshold $p^*$.

Notice that the definition of the $p$-value associated with a pair of
articles in terms of the hypergeometric null model provided in
Eq.~\ref{hyper} does not depend on the order in which two articles are
assessed. The resulting symmetric value of similarity between any two
papers is rooted in the invariance of the hypergeometric distribution
in Eq.~\ref{hyper} under permutation of the pair $i$ and $j$, i.e., of
the two quantities $d_i, d_j$. Moreover, Eq.~\ref{hyper} rectifies
some of the problems of measures of similarity based on a bare count
of co-occurrences. In particular, two articles that share a small
number $N_{ij}^k$ of citations will be assigned a higher $p$-value
(i.e., a smaller statistical significance of their similarity) than
two articles sharing a large number of citations. This means that, for
instance, the $p$-value $q_{P3, P4}(2)$ associated with the pairs of
articles $(P3, P4)$ in Fig.~\ref{fig:fig1}(b) will be smaller than the
$p$-value $q_{P5,P6}(2)$ associated with the pair of articles $(P5,
P6)$ in Fig.~\ref{fig:fig1}(c), since $P3$ and $P4$ share a larger
number of references (namely, four instead of two) to other articles
each receiving two citations. Moreover, the $p$-value associated with
the pair $(P7, P8)$ will be larger (i.e., the similarity between the
pair is less statistically significant) than the $p$-value associated
with the pair $(P9, P10)$. The reason lies in the fact that, according
to the hypergeometric null-model, the co-occurrence of a reference to
a highly-cited article is more likely to take place by chance than the
co-occurrence of a reference to an article with a relatively small
number of citations.

\begin{figure}[!h]
  \begin{center}
    \includegraphics[width=3in]{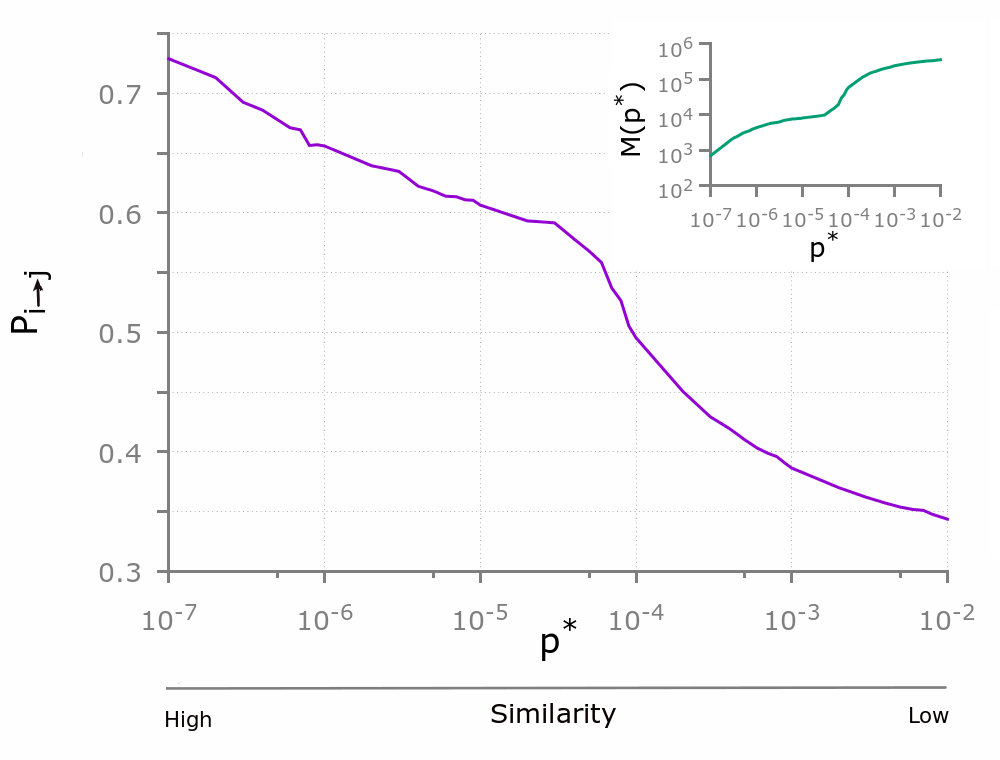}
  \end{center}
  \caption{The probability $P_{i\to j}(p^*)$ to observe a citation
    between two articles whose bibliographies overlap is statistically
    significant at the threshold value $p^*$. Notice that $P_{i\to
      j}(p^*)$ increases as the statistical threshold $p^*$
    decreases. That is, citations between pairs of articles
    characterised by a highly significant overlap tend to occur with a
    higher likelihood than citations between articles whose reference
    lists are not significantly similar. The inset shows how the
    number of pairs of articles characterised by a statistically
    significant similarity at a given threshold $p^*$ varies with
    $p^*$.}
  \label{fig:fig2}
\end{figure}

\section{Results}
\label{Results}

We now show how the proposed method for assigning a statistical
significance level to the similarity between any pair of articles
based on the statistically validated overlap between the respective
bibliographies can indeed turn very useful and help uncover important
properties of a citation network.

As an example of the possible applications of the method, we analyse
the citation network among articles published in the journals of the
APS during the period between 1893 and 2009. The data set is described
in detail in Section~\ref{Materials and Methods}.  We shall start by
studying empirically the probability $P_{i\to j}(p^*)$ of the
occurrence of a citation from an article $i$ to an article $j$
validated at a certain statistical threshold $p^*$.  We shall then
discuss how the method can be used to identify missing and potentially
relevant references and also to rank journals and scientific topics
based on the relative occurrence of missing citations.

\subsection{Homophily in citation patterns}

We start from the observation that if we consider progressively
smaller values of the statistical threshold $p^*$, the set
$\mathcal{M}(p^*)$ will shrink and contain only pairs of articles
characterised by an overlap between bibliographies that is highly
significant, since it has passed a more stringent statistical
test. Thus, small values of $p^*$ single out pairs of articles that
have a highly significant combination of common cited articles. But if
two articles share significantly similar bibliographies, then there is
a high probability that they are concerned with the same topic or
research problem. As a result, it would be reasonable to expect a
citation to occur from the more recently published article to the one
published at an earlier date.
For each value of the statistical threshold $p^*$, we computed the
number of pairs of articles $M(p^*)$ validated at that threshold in
the APS citation network, and the number $K(p^*)$ of existing
citations between those validated pairs. Then, we define the
probability $P_{i\to j}(p^*)$ that there exists a citation between any
two articles whose similarity is validated at the threshold $p^*$ as:
\begin{equation}
  P_{i\to j}(p^*) = \frac{K(p^*)}{M(p^*)}.
\end{equation}

The obtained values of $P_{i\to j}(p^*)$ are reported in
Fig.~\ref{fig:fig2} as a function of $p^*$.  The plot clearly suggests
that the probability of finding a citation between two articles
characterised by a highly statistically significant overlap between
the respective reference lists (i.e., the similarity between that pair
of articles is validated at a small value of $p^*$) is higher than the
probability of finding a citation between articles whose reference
lists are only moderately significantly similar. For instance, a
citation between a pair of articles $(i,j)$ whose overlap between
reference lists is validated at $p^*=10^{-2}$ occurs only with
probability $P_{i\to j}\simeq 0.35$, while citations occur within up
to $73\%$ of the pairs of articles validated at $p^*=10^{-7}$. In
other words, the probability that an article $i$ cites another article
$j$ is an increasing function of the similarity between the two
articles.

In the social sciences, the principle that similarity breeds
connection is traditionally referred to as homophily. This principle
has been documented in a variety of empirical domains
\cite{aral,kossinets,lazerfeld,mcpherson2001birds}. It is interesting
to observe that homophily can also be found to govern citation
networks where it plays an important role in shaping the structure and
evolution of knowledge transfer between academic papers.

\subsection{Suggesting missing references}

The identification of a statistically significant similarity between
two articles can be used to uncover potentially missing
references. For instance, the implementation of a recommendation
procedure based on statistically significant overlaps between
bibliographies might be useful to assist the editor of a scientific
journal in suggesting a list of possibly relevant (and missing)
references to the authors of a submitted paper.

\begin{figure*}
  \begin{center}
    \includegraphics[width=6in]{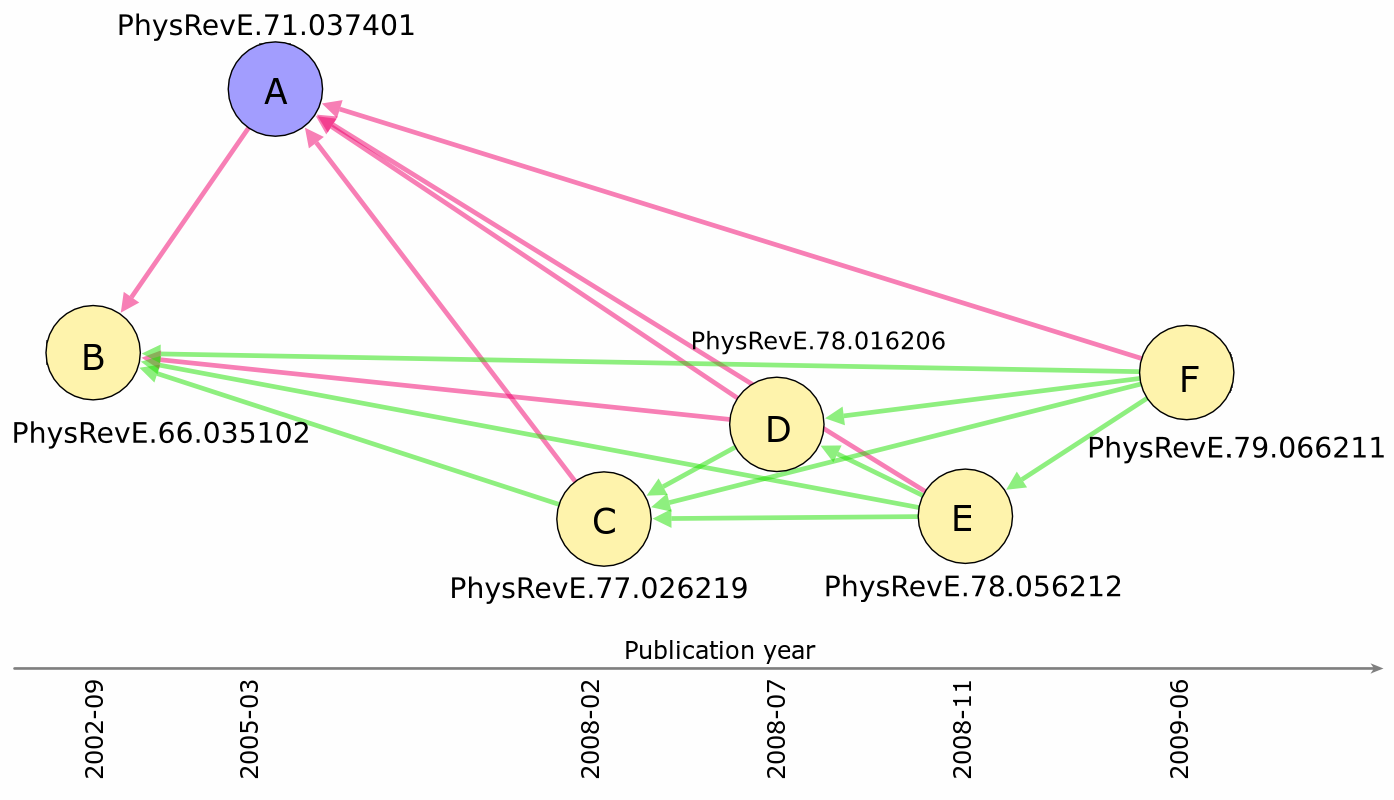}
  \end{center}
  \caption{\textbf{Lack of knowledge flows.} An example of several
    validated pairs of articles in the APS citation network at
    $p^*=10^{-7}$ (articles are reported in increasing order of
    publication time, from left to right). The occurrence of a link
    indicates that the pair of articles has passed the statistical
    test, while the colour of the link indicates that the most recent
    paper in the pair actually did (green) or did not (red) cite the
    other one. In this case, all the articles represented as yellow
    nodes are articles co-authored by researchers in the same group,
    while article A was co-authored by another group. The
    identification of a large number of missing citations suggests
    that the two groups might have been unaware of the work of their
    colleagues in the same field.}
  \label{fig:fig3}
\end{figure*}

Fig.~\ref{fig:fig3} shows a typical problem that could be fruitfully
addressed through an appropriate reference recommendation system based
on the identification of statistically significant overlaps between
bibliographies of papers. We report a subgraph of the APS citation
network consisting of several pairs of articles validated at
$p^{*}=10^{-7}$. Each article is represented as a node, and validated
pairs of nodes are connected through a link. The color of each link
indicates whether the older article was (green) or was not (red) cited
by the more recent one. Note that there is a prevalence of green
links, which is consistent with the fact that, for a significance
level $p^*=10^{-7}$, a citation between a validated pair of articles
occurs in more than $73\%$ of the cases (see
Fig.~\ref{fig:fig2}). However, we notice that article A has a
considerable number of missing citations, resulting from the fact that
it was not cited by any of the four articles that were published after
its publication date and with which it shares a statistically
significant portion of its bibliography (namely, nodes C, D, E,
F). This could mean that either the authors of articles C-F were not
aware of the existence of article A, despite the substantial overlap
between their reference lists, or that article A was not particularly
relevant to the topics addressed in the other articles.

Surprisingly, a more in-depth analysis of the articles in
Fig.~\ref{fig:fig3} suggests that, not only did all of them appear in
the same journal (Physical Review E), but indeed they are all
concerned with the same topic (electric discharges) and share a
relatively large fraction of PACS codes (05.45.-a, 52.80.Hc). The high
degree of similarity between topics can also be easily inferred from
the abstracts and introductions of these articles. Interestingly, we
found that articles B-F (yellow nodes) were all co-authored by the
same research group $G_1$, while article A (the only blue node) was
the result of the work of a different research group $G_2$. The fact
that also article A does not cite article B suggests that the
researchers in group $G_1$ were likely to be unaware of the work
conducted by group $G_2$ in the same research field, and vice-versa.

In this particular case, the quantification of statistically
significant overlaps between bibliographies could have been used to
facilitate the flow of knowledge between different research
groups. For instance, the editor of Physical Review E or the selected
reviewers could have brought article B to the attention of the authors
of article A, and similarly, when articles C-F were submitted to the
same journal, the editor or the reviewers could have advised the
authors of group $G_2$ to include article A in the bibliographies of
their submitted papers.

\begin{figure*}[!ht]
  \begin{center}
    \includegraphics[width=6in]{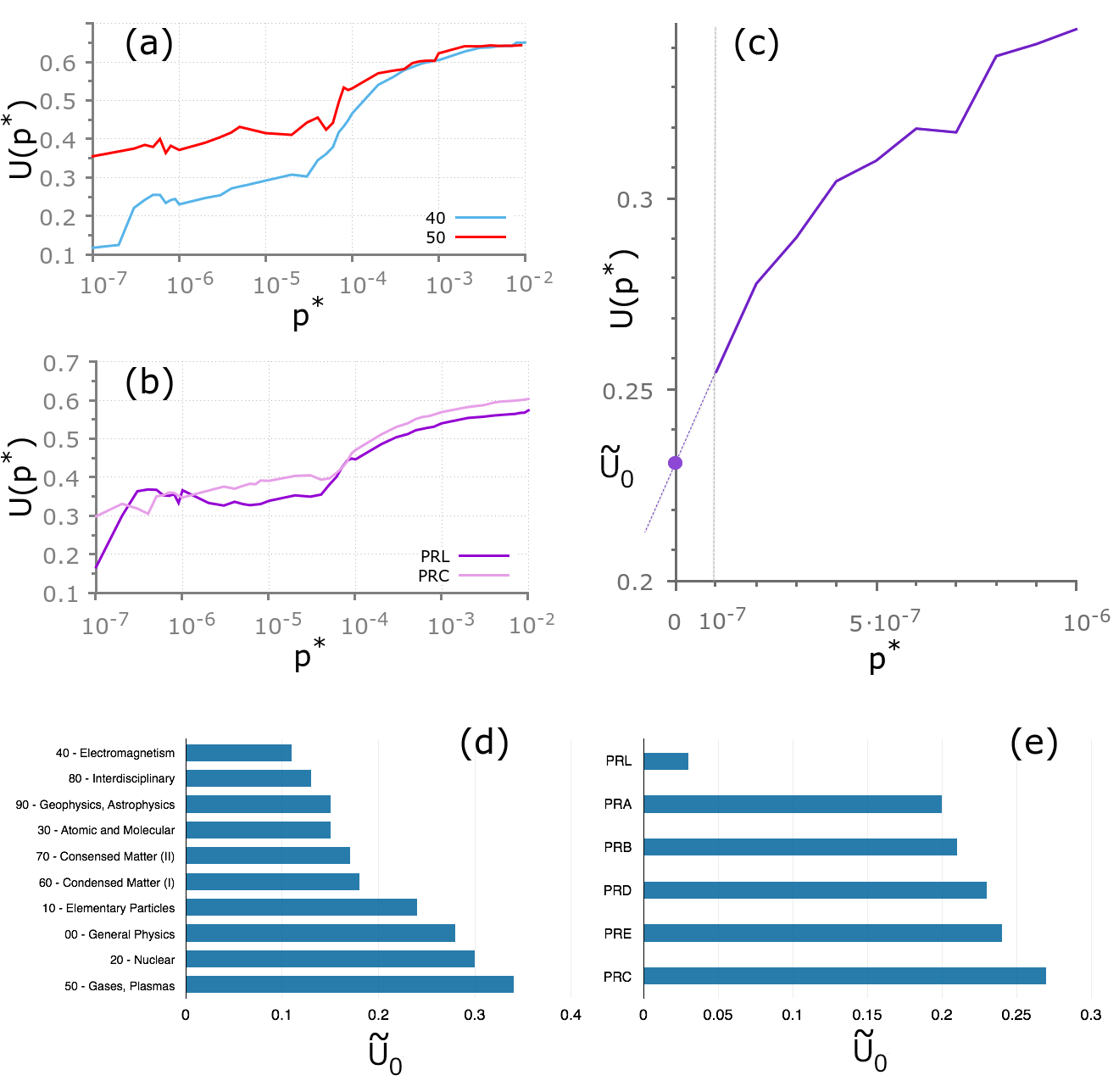}
  \end{center}
  \caption{\textbf{Ranking journals and sub-fields by lack of
      knowledge flows.} The analysis of missing links restricted to
    specific sub-fields of physics or single APS journals confirms
    that the tendency of a citation to occur between a pair of
    articles increases with the similarity between the bibliographies
    of the two articles. Panels (a)-(b) show the plots of $U(p^*) = 1
    - P_{i\to j}(p^*)$ for different sub-graphs corresponding to (a)
    two families of PACS codes, namely $40$ (electromagnetism) and
    $50$ (Gases and Plasmas), and (b) two APS journals, namely
    Physical Review Letters and Physical Review C. In panel (c) we
    sketch the procedure adopted to compute the estimate
    $\widetilde{U}_0$: we consider the line tangent to the curve
    $U(p^{*})$ at the smallest value of the statistical threshold
    $p^*$ for which we still have a relatively substantial number of
    validated pairs (in this case, $p^*=10^{-7}$), and we define
    $\widetilde{U}_0$ as the value of the intercept at $p^*=0$ of that
    line. In panels (d) and (e) we show, respectively, the rankings of
    sub-fields and APS journals based on the values of
    $\widetilde{U}_0$. Notice that Electromagnetism and
    Interdisciplinary physics are the two sub-fields with the smallest
    percentage of missing links, i.e., those in which knowledge among
    articles flows effectively and as would be expected if citations
    were driven by overlaps between topics or research
    problems. Interestingly, the lack of knowledge flows between
    articles published in Physical Review C ($\widetilde{U}_0\simeq
    0.27$) is almost nine times as large as the one identified in
    Physical Review Letters ($\widetilde{U}_0\simeq 0.03$), which is
    the APS journal with the widest visibility and largest impact.}
  \label{fig:fig4}
\end{figure*}

\subsection{Ranking journals and disciplines by (lack of) knowledge flows}

So far our analysis has been focused on the whole APS citation
network. Physics is a very broad disciplinary area, including
sub-fields as diverse as atomic physics, astronomy, particle physics,
statistical mechanics, just to mention a few
\cite{sinatra2015century}. It is therefore reasonable to perform our
analysis of the probability $P_{i\to j}(p^*)$ at the level of
sub-fields. Specifically, we argue that the percentage $P_{i\to
  j}(p^*)$ of citations occurring between pairs of articles associated
with a similarity that is validated at the statistical threshold $p^*$
can serve as a proxy for the knowledge flows taking place within a
sub-field. In what follows we restrict our analysis to the six
citation sub-graphs induced by the articles published in each of the
six research journals published by APS (in order to quantify the
ability of each journal to facilitate or impede the dissemination of
knowledge), and to the ten sub-graphs associated with the highest
levels in the PACS taxonomy (which could shed light on the typical
patterns of knowledge dissemination in different sub-fields).  The
lack of knowledge flows within a journal or a sub-field at a certain
confidence level $p^*$ can be quantified by the fraction of missing
links:

\begin{equation}
  U(p^*) = 1 - \frac{K(p^*)}{M(p^{*})} = 1 - P_{i\to j}(p^*).
\end{equation}

In general, the lower the value of $U(p^*)$, the more likely it is
that a citation occurs between a pair of articles characterised by a
similarity validated at the statistical threshold
$p^*$. Fig.~\ref{fig:fig4}(a)-(b) shows how $U(p^*)$ behaves as a
function of $p^*$, respectively, for all articles whose main PACS code
is either in group 40 (Electromagnetism) or in group 50 (Gases and
Plasmas), and for all the articles published in Physical Review
Letters and in Physical Review C. The figure clearly shows that, even
though in all cases $U(p^*)$ decreases when $p^*\to 0$, different
journals and different sub-fields tend to be characterised by slightly
different profiles of $U(p^*)$, namely by different propensities to
obstruct knowledge flows between similar academic papers. A
comparative assessment of journals and sub-fields according to their
typical ability to facilitate the dissemination of knowledge would, of
course, be based on $\frac{K(p^*)}{M(p^{*})}$. Moreover, the ranking
will in general depend on the chosen value of the statistical
threshold $p^*$.

From a theoretical point of view, a suitable approach to the ranking
would be to compute the quantity:

\begin{equation}
  U_{0} = \lim_{p^*\to 0} U(p^*),
\end{equation}
\noindent namely the limiting value of $U(p^*)$ when we let the
statistical threshold $p^*$ go to zero. However, this quantity cannot
be computed accurately for a finite network, since for a certain value
$p^*>0$ the number $M(p^*)$ of validated pairs at $p^*$ will be equal
to $0$, and the ratio $\frac{K(p^*)}{M(p^*)}$ would therefore be
undetermined. Here we employ a simple workaround, namely we consider
the tangent at the curve $U(p^*)$ at the smallest value of $p^*$ for
which the number of validated pairs is still large enough for the
construction of a network of a reasonable size (we found that
$10^{-7}$ is an appropriate choice in our case), and we compute the
intercept at which this tangent crosses the vertical axis. The value
obtained is denoted as $\widetilde{U}_{0}$, and is used as an
approximation of $U_{0}$. The procedure used to determine
$\widetilde{U}_{0}$ is sketched in Fig.~\ref{fig:fig4}(c).

In Fig.~\ref{fig:fig4}(d)-(e) we report the ranking induced by
$\widetilde{U}_{0}$ respectively for the ten high-level families of
PACS codes (panel d) and for the journals published by APS (panel
e). It is worth noticing that Electromagnetism and Interdisciplinary
Physics are the two sub-fields with the smallest percentage of missing
links, i.e., those in which knowledge flows effectively among articles
(and authors), as would be expected if the occurrence of citations
were driven by overlaps between topics or research
problems. Interestingly, the rate of occurrence of missing citations
in Physical Review C ($\widetilde{U}_0\simeq 0.27$) is almost nine
times as large as the one observed in Physical Review Letters
($\widetilde{U}_0\simeq 0.03$), which is the APS journal with the
widest visibility and largest impact.

\section{Conclusions}
\label{Conclusions}

In our study we have proposed a novel method for quantifying the
similarity between papers based on their bibliographies. The
identification of a statistically significant similarity between
papers can be used to uncover potentially interesting or relevant
references that are missing from their bibliographies. Our method can
thus assist the authors of scientific papers in compiling a list of
relevant references, or the editors and reviewers of scientific
journals in suggesting otherwise neglected references to the authors
of manuscripts submitted for publication. Moreover, public preprint
repositories, such as arXiv.org, could automatically quantify the
similarity between the bibliography of a newly posted paper and the
bibliographies of all other papers in their data set, and then propose
a list of papers that the authors might find relevant to their
work. The implementation of a recommendation procedure based on
statistically significant overlaps between bibliographies might also
facilitate the dissemination of scientific results within a scientific
field. Problems such as the one shown in Fig.~\ref{fig:fig3} can be
aptly overcome through the use of our method that enables missing and
relevant references to be promptly identified.

Since our analysis was based on the APS data set, the evaluation of
the similarity between any two articles was restricted to the overlap
between the citations the two papers made only to other papers
published in the APS journals. The assessment of similarity could not
therefore reflect the entire bibliographies of the two articles. This
limitation can be easily overcome through further analysis of other
citation networks extracted from different data sets, such as ISI Web
Of Science, or arXiv.org. Moreover, our framework can be extended
beyond the domain of citations between academic papers, and used for
uncovering missing and potentially relevant links in other citation
networks, such as those between patents
\cite{jaffe2002patents,sternitzke2008patent} or between the US Supreme
Court verdicts
\cite{clough2015transitive,fowler2008authority,fowler2007network}.

\section{Materials and Methods}
\label{Materials and Methods}
\subsection{The APS data set}

The APS data set includes bibliographic information on all the
articles published by the American Physical Society between 1893 and
2009 \cite{aps}. The citation graph $G=(V,E)$ includes $|V| = 450,084$
articles, and $|E| = 4,710,547$ directed links. The citations refer
only to articles that have been published on APS journals. For each
article we extracted the publication date, the main research subject
(according to the PACS taxonomy), and its bibliography. Each article
belongs to a specific journal. We restrict the analysis to the seven
major journals, namely Physics Review A, B, C, D, E and Letter, which
are specialised in different sub-fields of physics.

We performed our analysis at three levels, namely the entire
citation network, the sub-graphs of the citation network induced by
articles in each of the ten main sub-fields of physics, as identified
by the highest levels of the PACS hierarchy, and the six sub-graphs
induced by articles published in Physical Review Letters and in
Physical Review A-E. In our analysis, we discarded articles appeared in
Review of Modern Physics, which publishes almost exclusively review
articles. In Table~\ref{MacroPacs} we report the description of the ten
main categories in the PACS taxonomy and the topics covered by each of
the six journals here considered.

\begin{table*}[h!]
\caption{The scientific domains associated with the PACS codes and journals}
 	\begin{tabularx}{\textwidth}{c|X}
        \hline
       		\\
        \textbf{PACS code} &  \textbf{Domain} \\
		\\        
        \hline
        \\
        00 & General \\
        10 & The Physics of Elementary Particles and Fields\\
        20 & Nuclear Physics\\
        30 & Atomic and Molecular Physics\\
        40 & Electromagnetism, Optics, Acoustics, Heat Transfer, Classical Mechanics, and Fluid Dynamics \\
        50 & Physics of Gases, Plasmas, and Electric Discharges\\
        60 & Condensed Matter: Structural, Mechanical and Thermal Properties\\
        70 & Condensed Matter: Electronic Structure, Electrical, Magnetic, and Optical Properties \\
        80 & Interdisciplinary Physics and Related Areas of Science and Technology\\
        90 & Geophysics, Astronomy, and Astrophysics\\
        \\
        \hline
        \\
        \textbf{Journal} &  \textbf{Domain}\\
		\\        
        \hline
        \\
        Physics Review A & Atomic, molecular, and optical physics\\ 
        Physics Review B & Condensed matter and materials physics\\ 
        Physics Review C & Nuclear physics\\ 
        Physics Review D & Particles, fields, gravitation, and cosmology\\ 
        Physics Review E & Statistical, non-linear, and soft matter physics\\
        Physics Review Letter & Moving physics forward\\
		\\       
        \hline 
      \end{tabularx}
      \label{MacroPacs}
\end{table*}

\subsection{False Discovery Rate (FDR) statistical test}

The validation of a given pair $(i,j)$ in the FDR method is performed
as follows~\cite{Benjamini}. We set a statistical threshold $p^*$ and
we assume that there are in total $N_t$ tests.  Then, the $p$-values
of different tests are first arranged in increasing order $(q_1 < q_2
<...< q_{N_t})$, and the rescaled threshold is obtained by finding the
largest $t_{max}$ such that
\begin{equation}
q_{t_{max}} < \frac{p^* t_{max}}{N_t},
\end{equation}
\noindent where $N_t$ is the number of tests. In this specific case,
$N_t$ is the number of distinct pairs of papers that are tested over
all the sets $S^{k}$ of in-degree classes in the citation
network. Then we compare each $p$-value $q_{ij}(k)$ with the rescaled
threshold, and we validate the pair $(i,j)$ if $q_{ij}(k) < p^*
\,t_{max} /N_t$.

\begin{acknowledgments} 
  V.N. and V.L. acknowledge support from the Project LASAGNE, Contract
  No.318132 (STREP), funded by the European
  Commission. V.L. aknowledges support from EPSRC project GALE, Grant
  EP/K020633/1.
\end{acknowledgments}

\end{document}